\documentstyle[epsf,pstricks]{wasedapaper}
\def\asca	{{\em ASCA}\/}
\def\rosat	{{\em ROSAT}\/}
\def\am		{$^\prime$}

\def\muller	{M\"{u}ller}

\begin{document}

\title{ASCA ATLAS OF THE CLUSTER TEMPERATURES}

\author{Maxim L. Markevitch$^1$, Craig L. Sarazin$^1$ 
and Mark J. Henriksen$^2$\\
$^1${\em University of Virginia, Charlottesville, VA 22903, USA}\\ 
$^2${\em University of North Dakota, Grand Forks, ND 58202, USA}} 

\maketitle

\section*{Abstract}

We use \asca\ data to obtain two-dimensional maps of the gas temperature in
three clusters: A754, A3558 and Triangulum Australis (the maps of A2256,
A2319, A2163 and A665, also presented at the conference, have since appeared
in [5]). All clusters from our sample show considerable temperature decline
with radius at $r\sim 0.5-1\,h^{-1}$ Mpc, most prominently in distant A2163
and A665. The three clusters presented here also feature asymmetric spatial
temperature variations, which may be naturally attributed to the effects of
a subcluster merger. As an example, we show that the Triangulum Australis gas
temperature and density maps indicate recent nonadiabatic heating,
presumably by merger shocks. Unlike most of the clusters, the systems in
our sample lack cooling flows (with the possible exception of a weak one in
A3558), thus we may be probing the younger members of the cluster
population.

\vspace{7mm}

Spatially resolved measurements of the gas temperature in clusters of
galaxies can provide valuable information on the dynamical history of these
systems, pointing to those clusters with recent or ongoing merger activity
(e.g., [8]). The present-day cluster merger rate is dependent on
cosmological parameters, with more mergers in higher-$\Omega$ models.
Simulations show that cluster radial temperature profiles are also sensitive
to the underlying cosmology (e.g., [2]). Thus, measurements of the cluster
temperature structure are of significant interest. Such measurements for hot
(and therefore massive) clusters have become possible now with the advent of
\asca.

\asca\ temperature maps of A2256, A2319, A2163 and A665 were presented in
[5] (see also [1]). In this paper, we present our results for A754, A3558
and Triangulum Australis (all three have $z\simeq 0.05$). To reconstruct
their temperatures, we employed the technique described in [5,6], using
\rosat\ images as a brightness template and modeling \asca\ mirror
scattering, modifying the technique for the cD region in A3558 [7].

Earlier, \rosat\ PSPC data suggested the existence of gas temperature
variations over the face of A754 [4]. The cluster galaxy distribution and
its X-ray image indicate the ongoing subcluster merger, and the temperature
nonuniformity supports this hypothesis. Our GIS+SIS results for A754,
presented in Fig.~1 (and in more detail in [3]), are in general agreement
with earlier findings while having a much better accuracy. We detect a spur
of hot gas along the cluster elongation axis. Cooler gas resides in the
north-eastern outskirts, and, according to [4], in the region of the
brightness peak. The temperature and brightness maps of A754 are strikingly
similar to the results of a simulation of a merger with the nonzero impact
parameter [2]. If the analogy with the simulation is correct, the eastern
and western subunits seen in the image are infalling from North and South,
respectively. The elongated plume of cool gas near the X-ray brightness peak
is then the stripped atmosphere, and perhaps even a cooling flow, belonged
to the eastern subunit.

\begin{figure}[t]
\pspicture(0,8.9)(14.5,20)
%\psgrid(0,8.9)(14.5,20)

\rput[tl]{0}(-0.4,19.8){\epsfysize=7.4cm
\epsffile{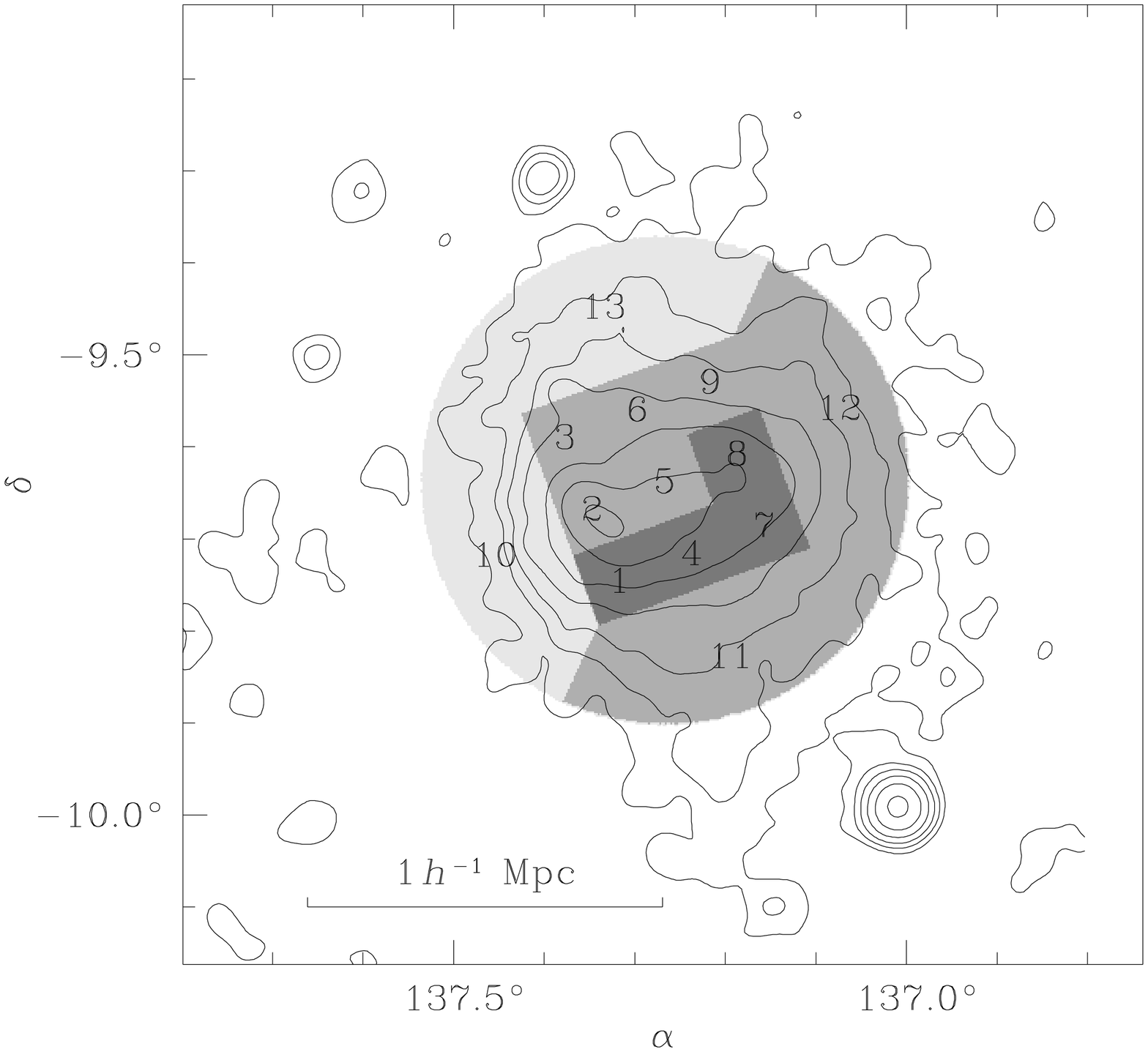}}

\rput{0}(-0.5,16.5){\psframebox*{\white \rule{5mm}{5mm}}}

\rput[tl]{0}(-0.35,12.8){\epsfxsize=8.5cm
\epsffile{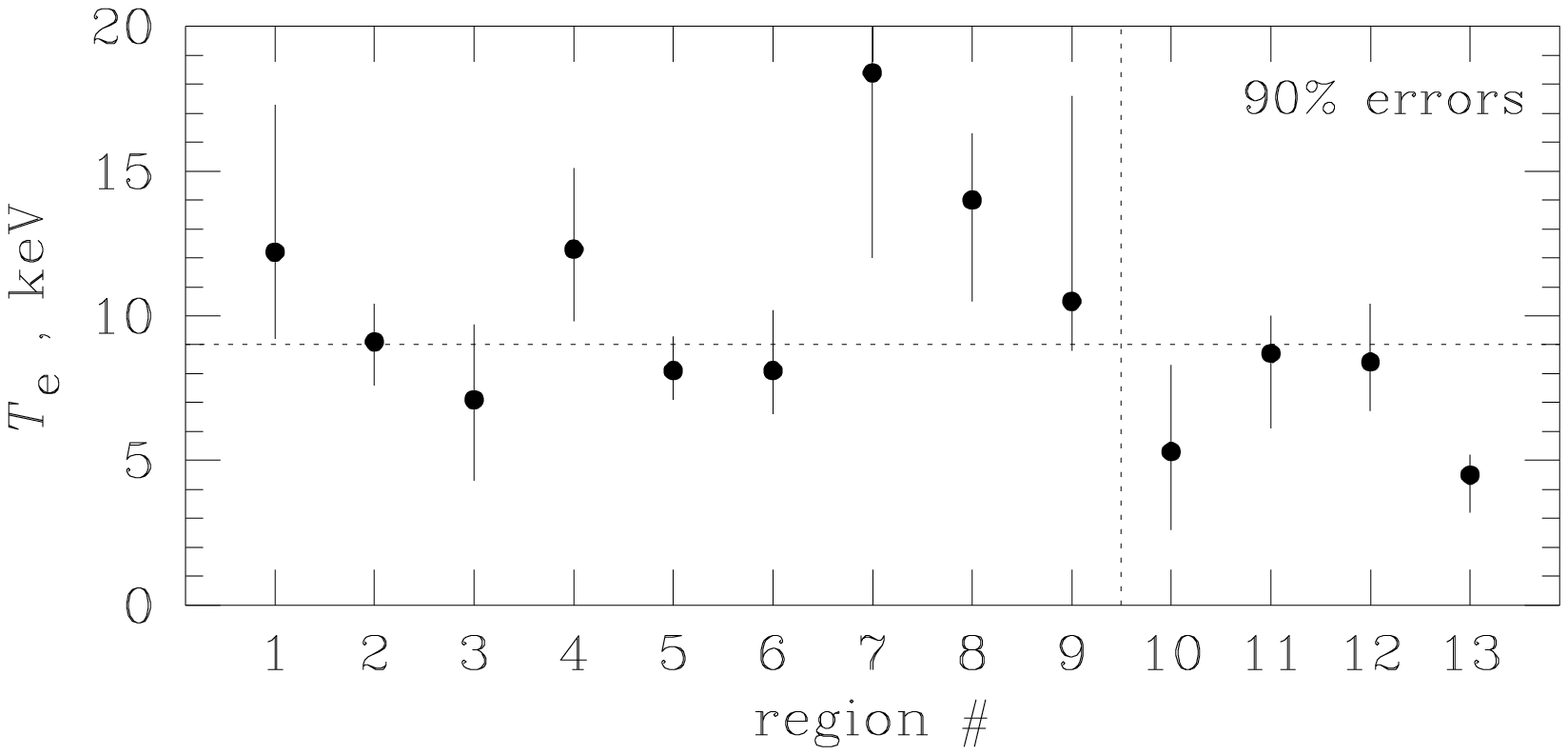}}

\rput[tl]{0}(7.1,19.8){\epsfysize=7.4cm
\epsffile{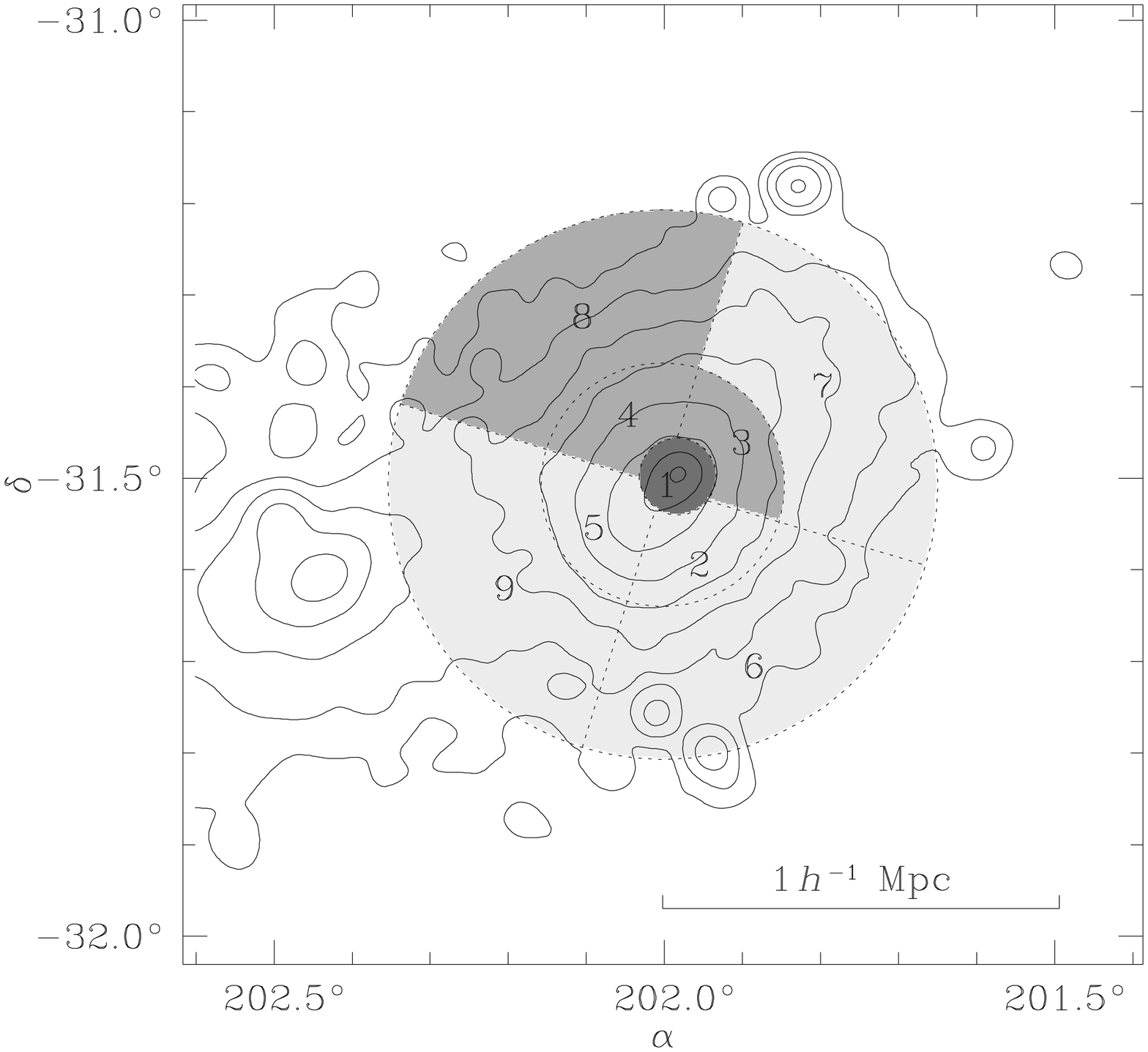}}

\rput[tl]{0}(7.3,12.8){\epsfxsize=7.5cm
\epsffile{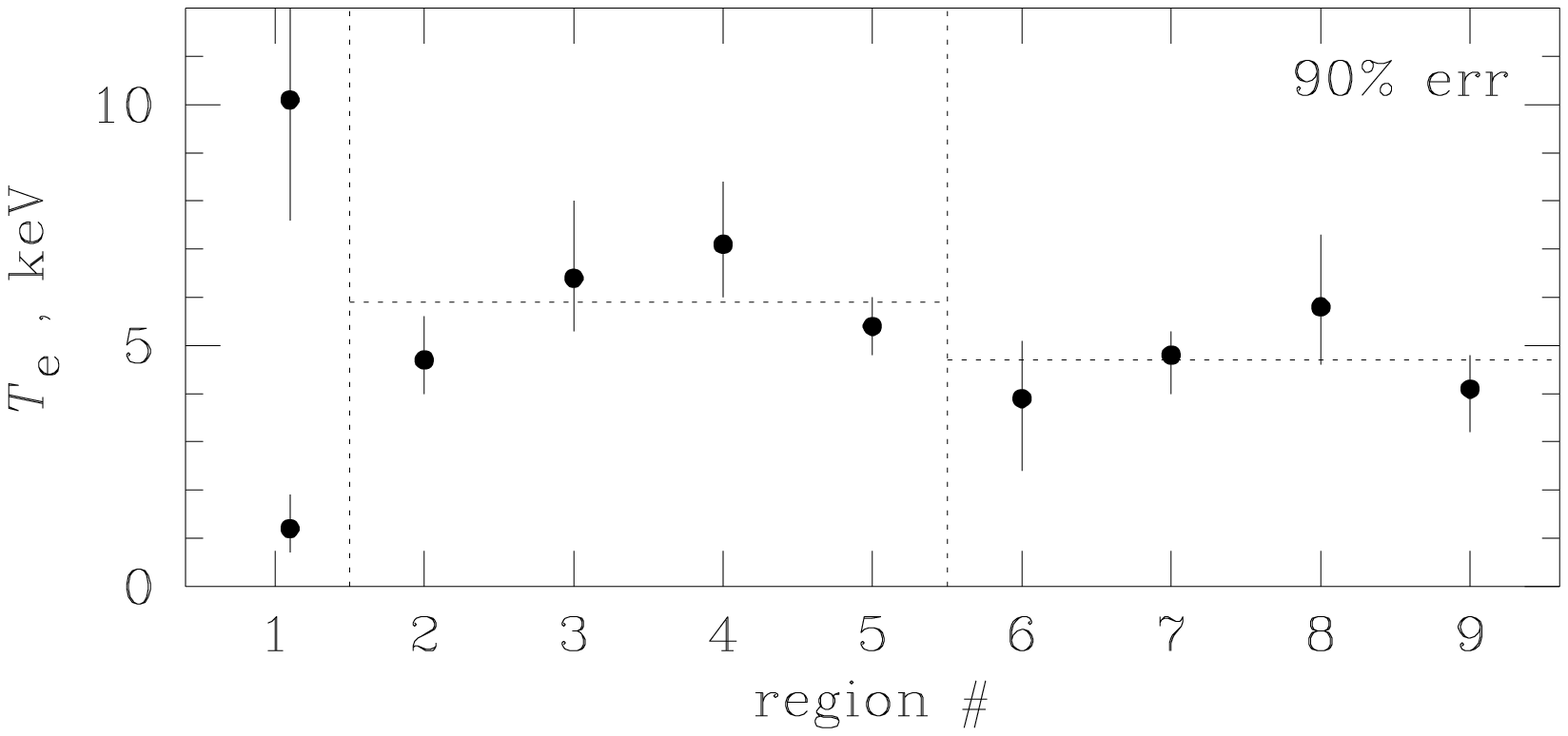}}

\rput{0}(3.9,19.85){A754}
\rput{0}(11.4,19.85){A3558}

\rput[cl]{0}(5.5,12.1){\psframebox*{\white \rule{12mm}{1mm}}}
\rput[cl]{0}(13,12.1){\psframebox*{\white \rule{11mm}{1mm}}}
\rput[cl]{0}(3.2,9.25){\psframebox*{\white \rule{11mm}{1mm}}}
\rput[cl]{0}(10.8,9.25){\psframebox*{\white \rule{11mm}{1mm}}}
\rput[cl]{90}(0.1,10.5){\psframebox*{\white \rule{11mm}{1.5mm}}}
\rput[cl]{90}(7.6,10.5){\psframebox*{\white \rule{11mm}{1.5mm}}}

\rput{0}(3.9,9.25){region \#}
\rput{0}(11.4,9.25){region \#}
\rput{90}(0.2,11){$T_e$, keV}
\rput{90}(7.7,11){$T_e$, keV}
\endpspicture

Fig.~1---Temperature maps of A754 and A3558. Contors show \rosat\ PSPC
surface brightness; grayscale shows \asca\ temperatures. For A754, regions
are 5\am\ boxes and sectors of the $r=16'$ circle. For A3558, the regions
are an $r=2.5'$ circle centered on the cD galaxy and sectors of the two
annuli $r=8-18'$ around the cluster centroid.  Regions are numbered and
their temperatures are shown in the lower panels.  Two spectral components
are shown for the cD region in A3558. Horizontal lines correspond to the
average temperature in A754 and average temperatures within each annulus for
A3558. All errors are 90\%.
\end{figure}

A3558 is a core member of the Shapley supercluster. Its X-ray image suggests
that the cluster may have experienced a merger and more are to happen. The
temperature map and radial profile of A3558 are shown in Figs.~1 and 3. A
fit to the cD galaxy region required at least two spectral components, with
the additional hot thermal component preferred over a physically meaningful
power law [7]. There is a slight radial temperature decline, which is
asymmetric. The asymmetry is present in both SIS and GIS data fitted
separately, although with marginal significance. We interpret this asymmetry
as a merger signature.

A \rosat\ image of the Triangulum Australis cluster shows that there is an
underdense region east of center (Fig.~2; region 2). From the image alone,
one expects this region to have a higher temperature to maintain a pressure
similar to the adjoining sectors, assuming the cluster gravitational
potential is reasonably smooth. Our temperature map indicates that the
cluster core has a higher temperature, and the temperature in the
low-brightness sector is indeed slightly higher than the average at that
radius, just enough to balance the gas pressure.  However, its specific
entropy, which is a useful diagnostic of shock heating or any other
nonadiabatic heating, is significantly higher than that in other regions at
the same radius (Fig.~2). The entropy in the core is also higher than that
expected in an isothermal model, which is a kind of distribution a cluster
gas is expected to approach if left to itself, particularly if thermal
conduction is effective. The simplest explanation of such pressure and
entropy distribution would be that the gas in the cluster center was heated
by the passage of a merger shock. The same event might have heated and
ejected the gas in region 2 from the center, which then adiabatically
expanded to its present density. Hydrodynamic simulations of a head-on
merger [8] predict a hot core and an asymmetric temperature structure not
unlike the observed one.

\begin{figure}[t]
\pspicture(0,8)(14.5,20)
%\psgrid(0,8)(14.5,20)

\rput[tl]{0}(-0.1,20.2){\epsfysize=7.5cm
\epsffile{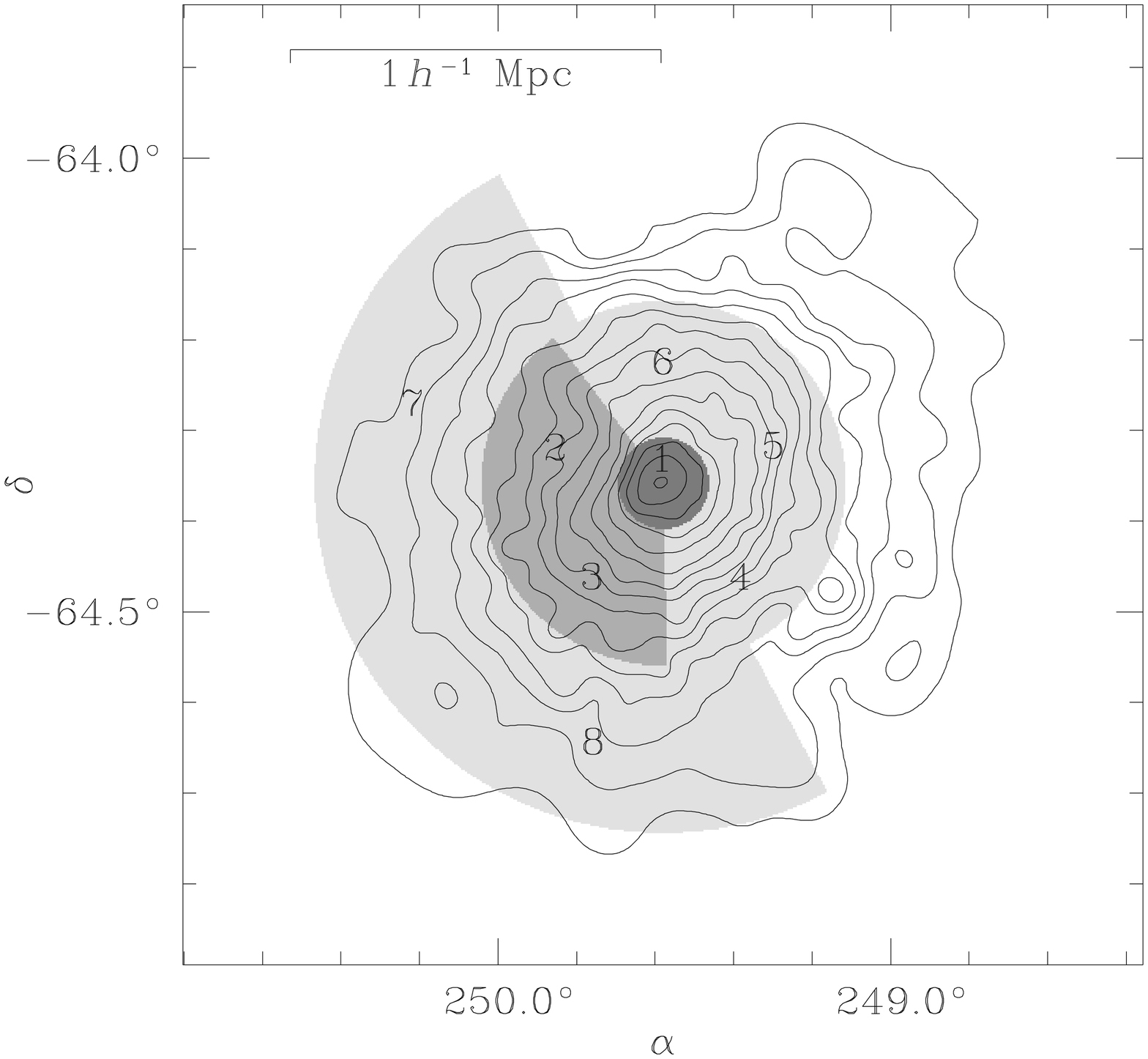}}

\rput[tl]{0}(7.6,20.13){\epsfxsize=7cm
\epsffile{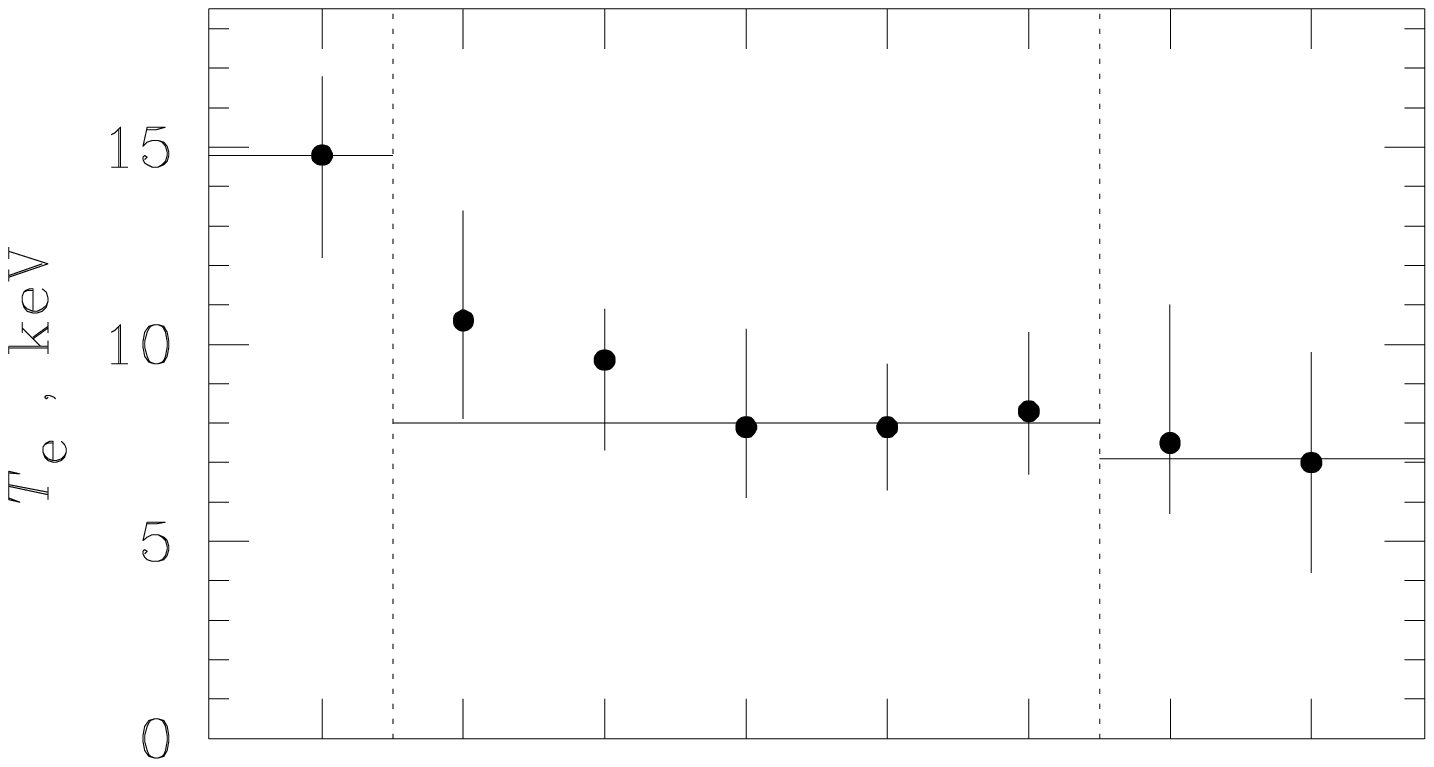}}

\rput[tl]{0}(7.6,16.3){\epsfxsize=7cm
\epsffile{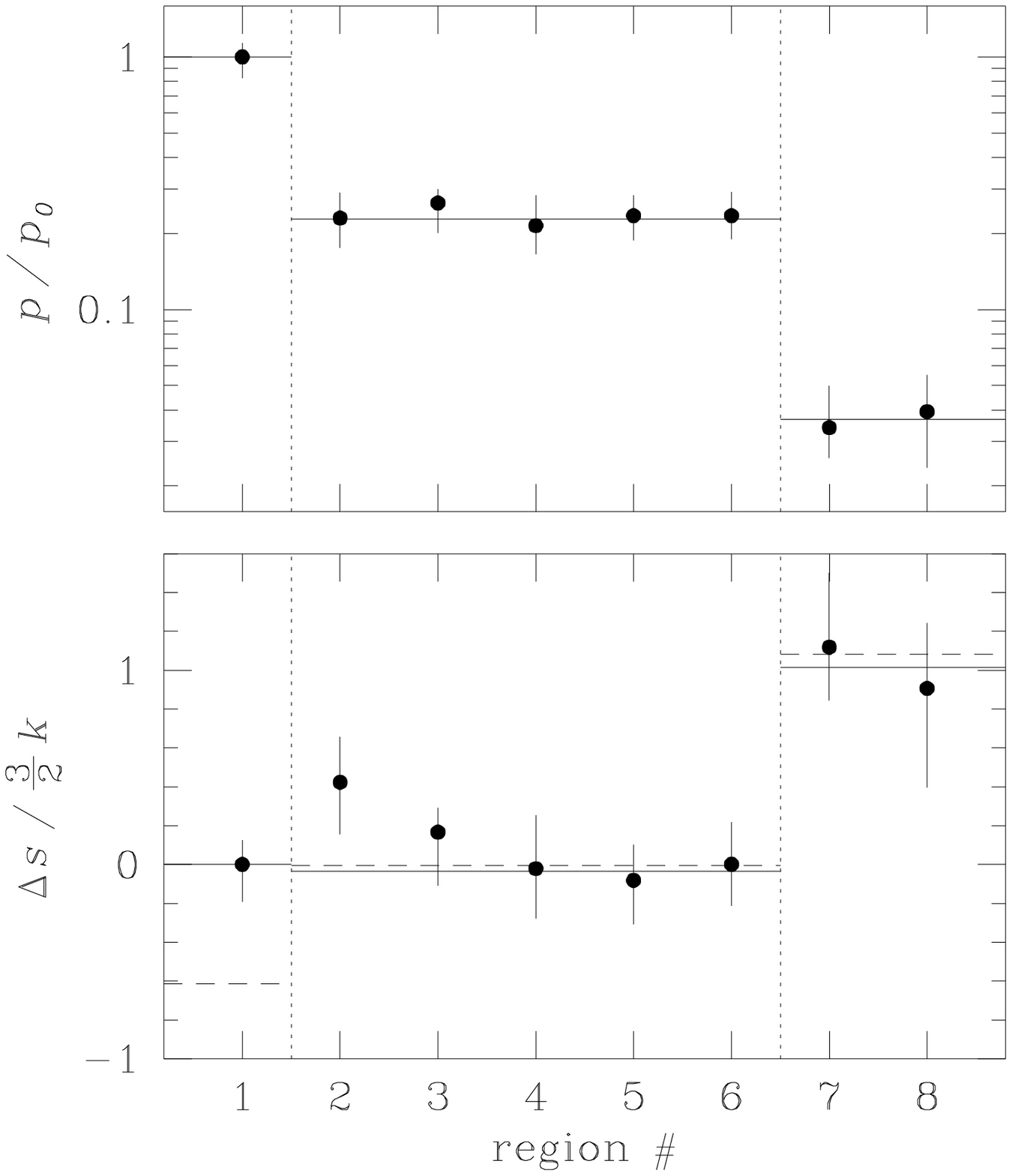}}

\rput[cl]{0}(10.9,8.5){\psframebox*{\white \rule{15mm}{2mm}}}
\rput{0}(11.5,8.4){region \#}

\rput[tl]{0}(-0.1,12.6){
\begin{minipage}{7.4cm}
Fig.~2---Temperature map of Triangulum Australis, overlaid on the \rosat\
image. \asca\ regions are concentric annuli with $r=3-12-23'$, with the
outer annuli divided into 5 and 4 sectors, respectively (two sectors poorly
covered by \asca\ are not shown.)  Right panels show temperature, gas
pressure and specific entropy in these regions.  Horizontal line in the
second annulus corresponds to an average over regions 4, 5 and 6. Dashed
line shows entropy for an isothermal $\beta$-model with $T_e=8$ keV.
\end{minipage}
}
\endpspicture
\end{figure}

\vspace{5mm}

Of the eight clusters for which the temperature was reconstructed so far
using our method, all but one exhibit a radial temperature drop at $r\sim
0.5-1\,h^{-1}$ Mpc (the exception is nearby AWM7, for which our analysis
covered only $r<0.25\,h^{-1}$ Mpc and resulted in a constant profile [7]).
Fig.~3 shows the radial temperature profiles of those clusters which are
relatively symmetric. Of the five clusters reasonably well resolved by
\asca\ (A2256, A2319, A754, A3558 and Triangulum Australis), three,
presented here, show asymmetric temperature structure suggestive of a recent
or ongoing merger. The other two, A2256 and A2319, although not showing the
characteristic temperature asymmetries [5], are probably starting to merge
with the subunits which are apparent in their X-ray images, but the process
may have not yet disturbed the bulk of gas.

An eventual drop of the gas temperature at some distance from the cluster
center is expected intuitively and is predicted by simulations (e.g., [2]).
However, in the clusters we studied, the temperature appears to decline even
steeper than expected in the standard cosmological models. The most extreme
case of A2163 is discussed in [6]. Generally, steeper temperature profiles
are predicted in either the models with low $\Omega$ (due to the steeper
density profiles), or in the models with significant galaxy feedback [2]. In
open models, present-day clusters are relaxed and there are few mergers
among them.

However, our cluster sample is quite likely to be biased towards younger
systems, because of our selection of objects without cooling flows to
minimize the effects of the uncertain \asca\ PSF. Cooling flows may be
disrupted by mergers, and merging clusters are expected to have peaked
temperature profiles [8]. As our sample expands and becomes more
representative, it may become possible to derive cosmological constraints
from it.

\begin{figure}[t]
\pspicture(0,6.6)(14.5,20)
%\psgrid(0,6.6)(14.5,20)

\rput[tl]{0}(0.4,20.4){\epsfysize=7cm
\epsffile{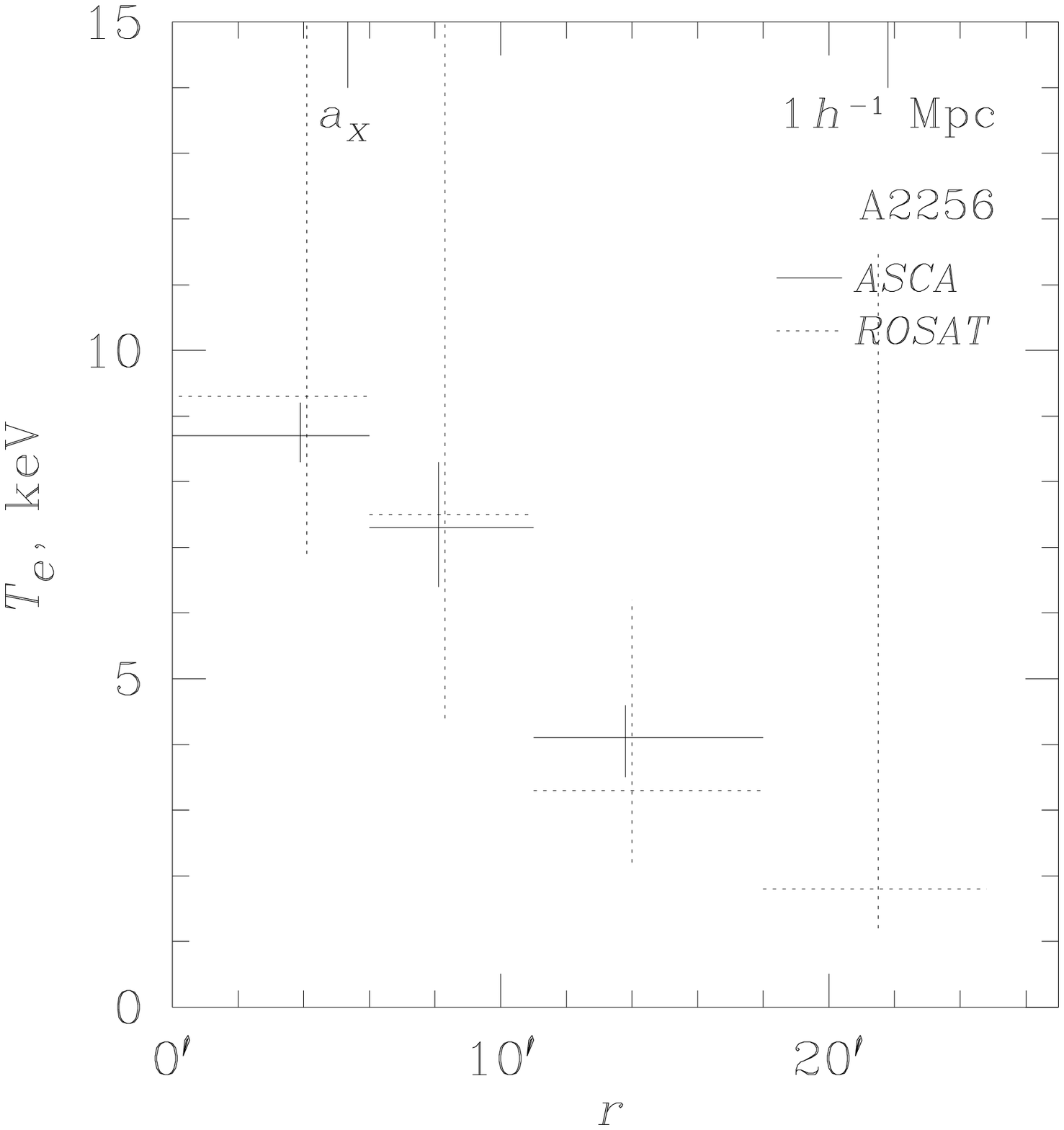}}

\rput[tl]{0}(7.1,20.4){\epsfysize=7cm
\epsffile{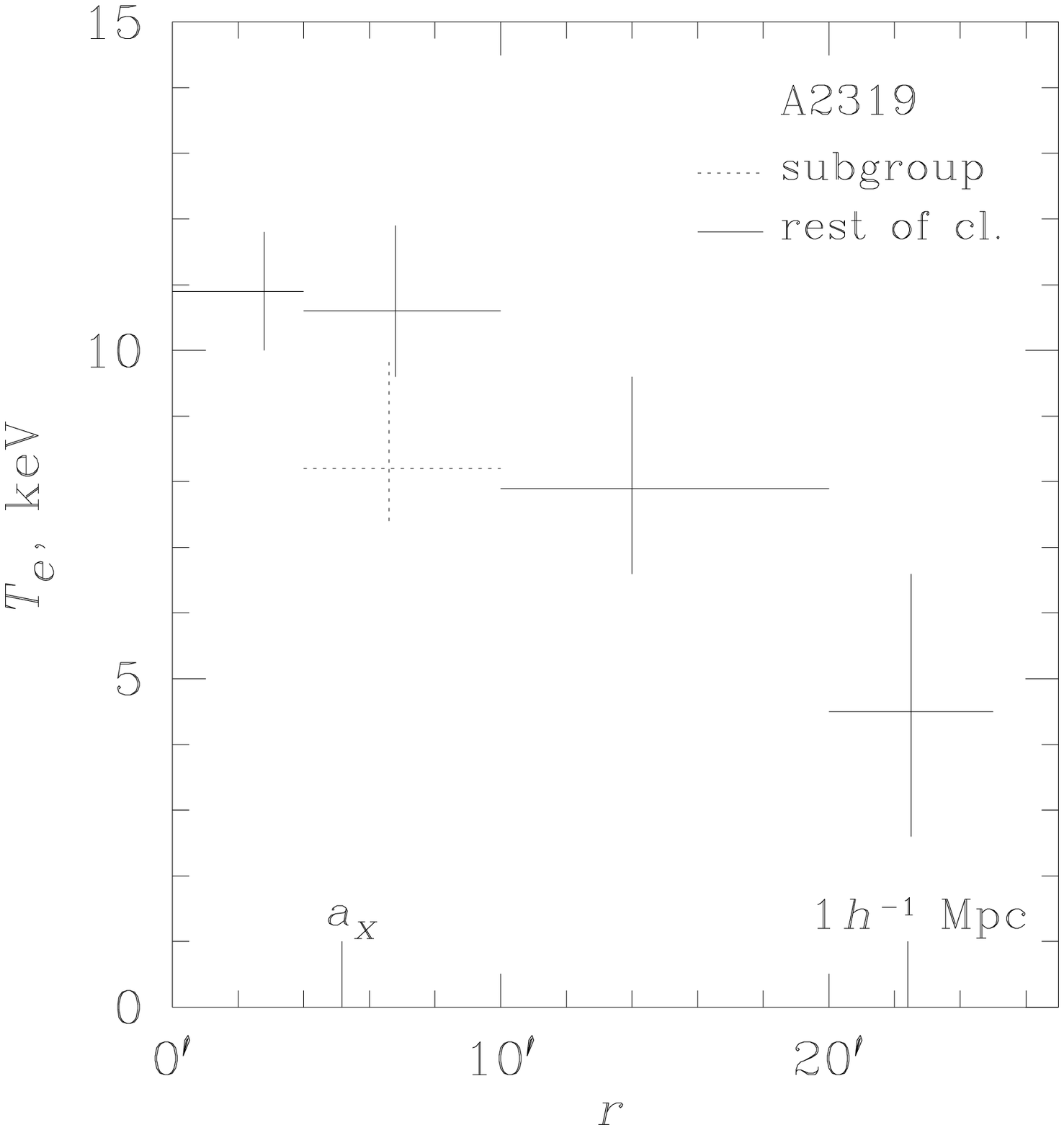}}

\rput[tl]{0}(0,13.6){\epsfysize=7cm
\epsffile{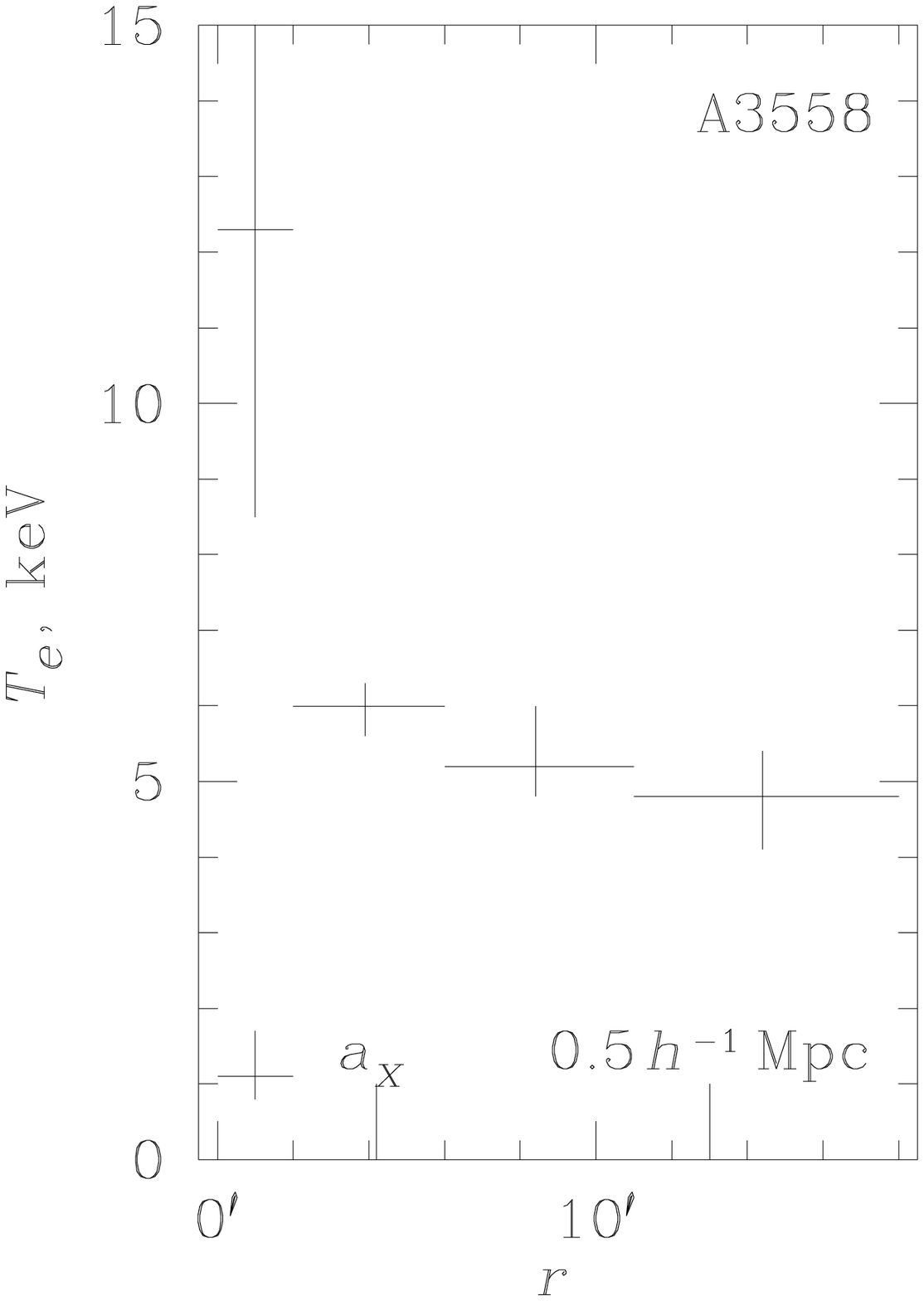}}

\rput[tl]{0}(5.75,13.6){\epsfysize=7cm
\epsffile{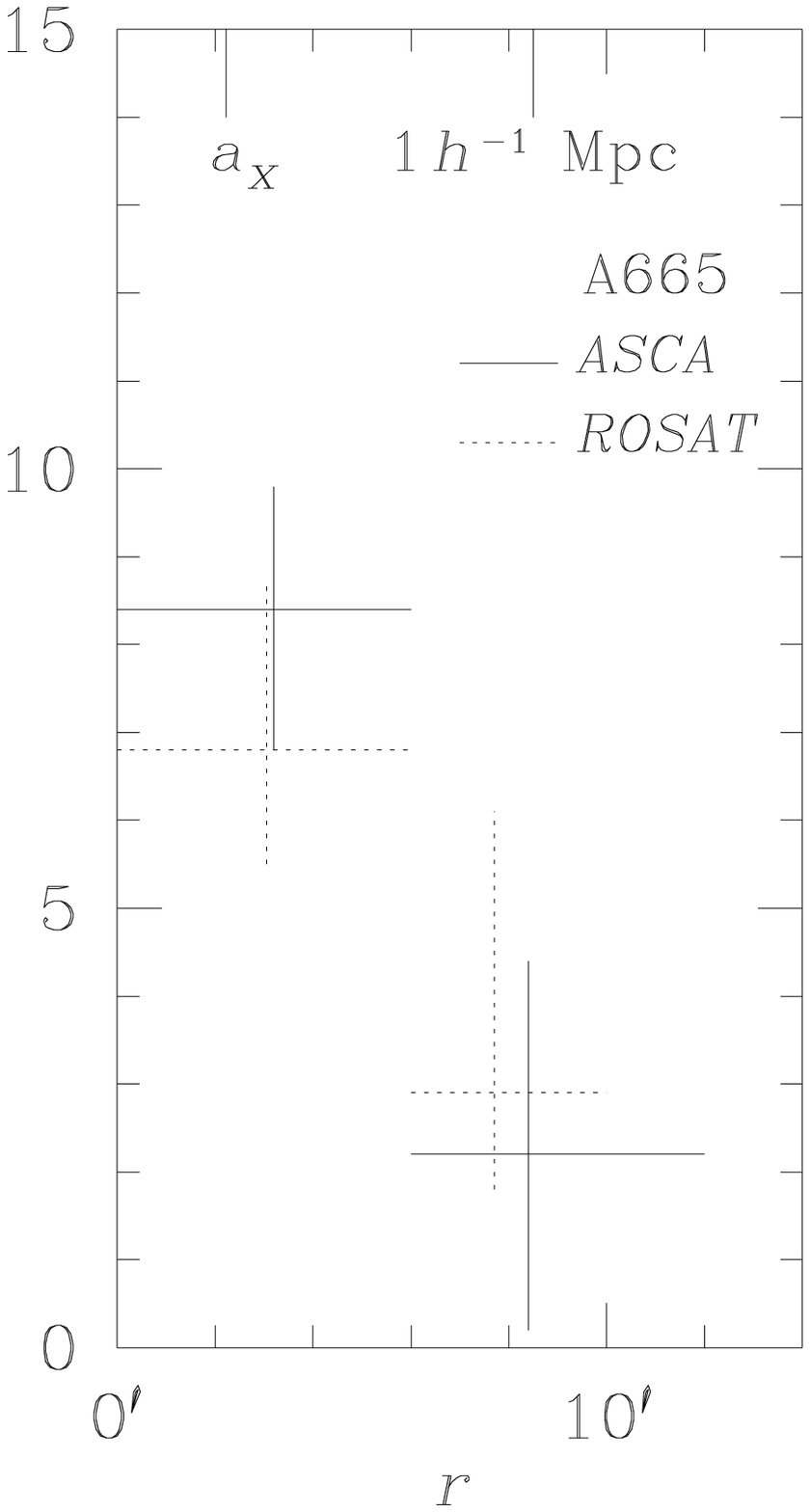}}

\rput[tl]{0}(10.2,13.6){\epsfysize=7cm
\epsffile{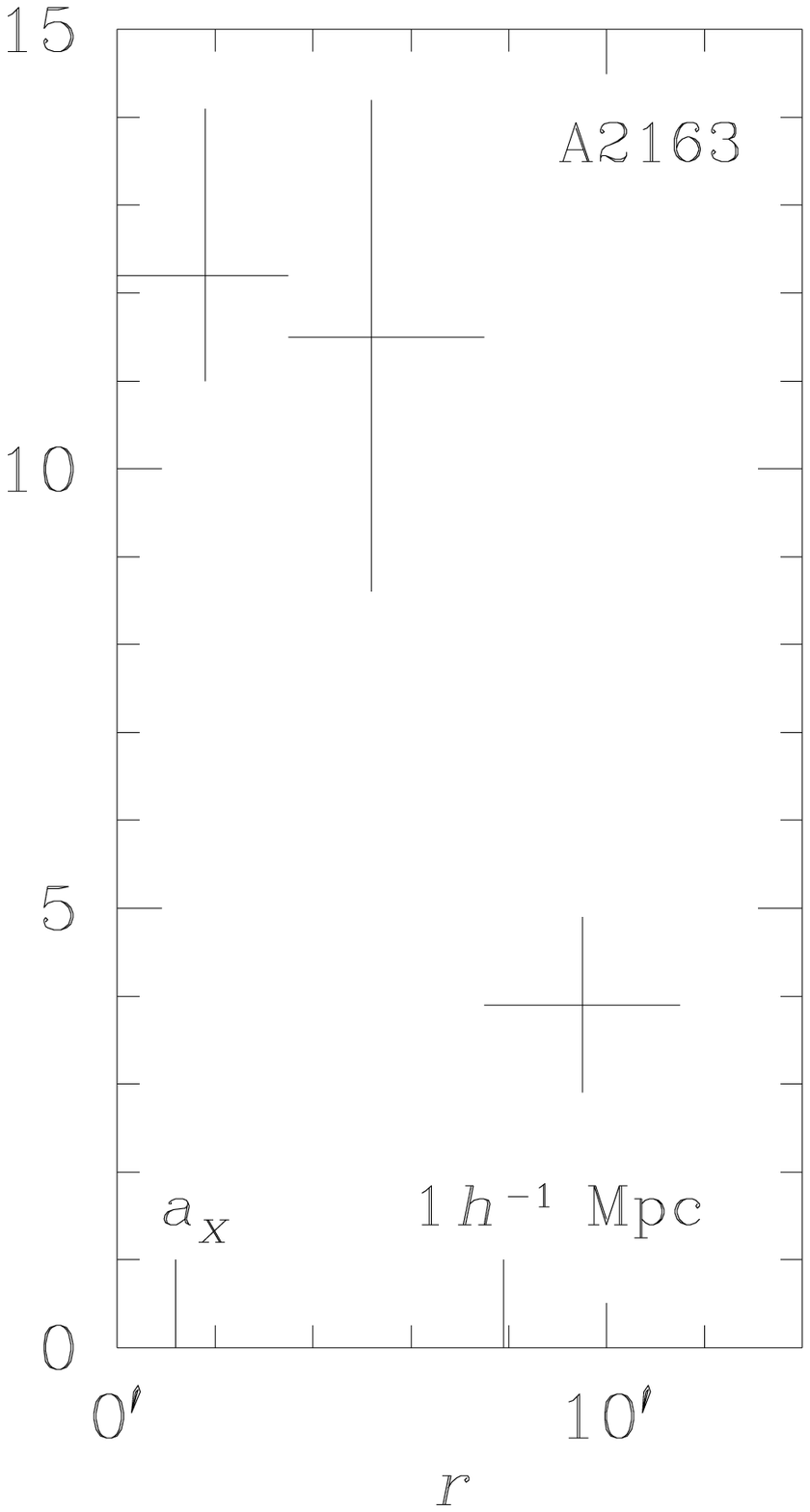}}
\endpspicture

Fig.~3---Radial temperature profiles of five clusters. For A2163,
temperatures in three-dimensional shells are shown (from [6]), while other
temperatures are projected on the image plane. For A2256 and A665, also
shown are \rosat\ PSPC results from [5,7].
\end{figure}

\vspace{1pc}

\re 
1. Churazov, E., Gilfanov, M., Forman, W., \& Jones, C. 1996, these proceedings
\re
2. Evrard, A. E., Metzler, C. A., \& Navarro, J. F. 1996, ApJ in
press; astro-ph/9510058
\re
3. Henriksen, M. J. \& Markevitch, M. 1996, ApJ Lett.\ in press;
astro-ph/9604150
\re
4. Henry, J. P., \& Briel, U. 1995, ApJ, 443, L9
\re 
5. Markevitch, M. 1996, ApJ, 465, L1
\re
6. Markevitch, M., et al.\ 1996, ApJ, 456, 437
\re
7. Markevitch, M. \& Vikhlinin, A. 1996, ApJ submitted; astro-ph/9605026
\re
8. Schindler, S., \& \muller, E. 1993, A\&A, 272, 137

\end{document}